\documentclass[sigconf]{acmart}

\AtBeginDocument{%
  \providecommand\BibTeX{{%
    \normalfont B\kern-0.5em{\scshape i\kern-0.25em b}\kern-0.8em\TeX}}}

\copyrightyear{2024}
\acmYear{2024}
\setcopyright{cc}
\setcctype{by}
\acmConference[CHI PLAY Companion '24]{Companion Proceedings of the Annual Symposium on Computer-Human Interaction in Play}{October 14--17, 2024}{Tampere, Finland}
\acmBooktitle{Companion Proceedings of the Annual Symposium on Computer-Human Interaction in Play (CHI PLAY Companion '24), October 14--17, 2024, Tampere, Finland}\acmDOI{10.1145/3665463.3678819}
\acmISBN{979-8-4007-0692-9/24/10}

\usepackage{comment}

\begin{document}

\title[Refresher Training Game for ASHAs and AWWs in India]
{Refresher Training through Digital and Physical, Card-Based Game for Accredited Social Health Activists (ASHAs) and Anganwadi Workers (AWWs) in India}

\author{Arka Majhi}
\email{arka.majhi@iitb.ac.in}
\affiliation{
  \institution{Indian Institute of Technology Bombay}
  \streetaddress{Powai}
  \city{Mumbai}
  \state{Maharashtra}
  \country{India}
  \postcode{400076}
}
\orcid{https://orcid.org/0000-0002-5057-1878}

\author{Aparajita Mondal}
\email{aparajita.mondal@tuni.fi}
\affiliation{
  \institution{Tampere University}
  \streetaddress{Kalevantie 4}
  \city{Tampere}
  \state{}
  \country{Finland}
  \postcode{33100}
}
\orcid{https://orcid.org/0000-0003-4609-2249}

\author{Satish B. Agnihotri}
\email{sbagnihotri@iitb.ac.in}
\affiliation{
  \institution{Indian Institute of Technology Bombay}
  \streetaddress{Powai}
  \city{Mumbai}
  \state{Maharashtra}
  \country{India}
  \postcode{400076}
}
\orcid{https://orcid.org/0000-0002-0703-3185}

\renewcommand{\shortauthors}{Arka Majhi, Aparajita Mondal, \& Satish B. Agnihotri}

\begin{abstract}

India's recent health surveys have highlighted a worrying trend of incomplete child immunization rates across several district clusters in India. Conventional training methods for community healthcare workers (CHWs) in India are inadequate for improving their skills and knowledge. Smartphone games could be a viable and cost-effective method of refresher training specifically targeting immunization practices. A refresher training game was designed both as a physical card-based and digital app-based game, focusing on enhancing CHWs' knowledge and practices related to child immunization. A quasi-experimental study was conducted with 368 participants. Quantitative gameplay analytics and qualitative feedback from players were collected through interviews. The findings show that game-based refresher training significantly improves CHWs' knowledge gain and retention in the area of child immunization. The discussion highlights the study's implications and insights while developing effective digital tools for training CHWs. The research contributes to the growing body of work on digital tools for training CHWs in resource-constrained settings. The study underscores the potential of smartphone games as a scalable and effective method of refresher training for improving child immunization rates.

\end{abstract}




\maketitle

\section{Introduction}

Community Health Workers (CHWs) are crucial in improving health outcomes in resource-constrained settings by improving immunization coverage in India and providing information, education, and vaccination services to the community. However, due to inadequate training, supervision, and resources, CHWs face several challenges in providing effective immunization services. Furthermore, CHWs often have limited literacy and smartphone usage skills, which can hinder their ability to provide effective healthcare services. Maintaining the quality of CHWs' work requires ongoing training and support \cite{Mishra2024}. Refresher training is crucial for CHWs to stay updated on immunization guidelines and practices. Traditional methods of training, such as classroom-based lectures in teaching topics like immunization, have been found to be ineffective in improving CHWs' knowledge and skills \cite{Patel2010}. They are time-consuming and costly and may not be scalable enough to reach many CHWs across the country. Therefore, there is a need for innovative and engaging training methods that can effectively reinforce knowledge and skills among CHWs \cite{Yadav2021}. During a dialogue session between the Women in Global Health (WGH) India chapter and Accredited Social Health Activists (ASHA) workers, it was suggested to implement a capacity building strategy utilizing technology and supervision mechanisms \cite{Asthana2022}.

Smartphone games are emerging as a scalable and effective way to provide refresher training to CHWs in India and other low and middle income countries \cite{Agarwal2016}. These games can be designed to be engaging and interactive, tailored to the needs and preferences of CHWs, and can be accessed remotely. Furthermore, they can be designed to be compatible with low literacy and smartphone usage skills, ensuring that they are accessible to a broad range of CHWs. While digital and smartphone-based games have gained popularity in recent years, physical card games remain an attractive option for the refresher training of CHWs. Digital games require access to smartphones or other devices, which may not be available or affordable for all CHWs. In addition, digital games can face technical challenges such as poor internet connectivity, low battery life, and limited storage space. On the contrary, physical card games require only cards that can be easily transported and used in low-resource settings. However, digital games have some advantages over physical card games. Digital games can provide instant feedback to players, and game analytics can be used to track player progress and improve the effectiveness of training. Digital games can also provide a more immersive and interactive experience for players, incorporating multimedia elements such as videos and animations. Therefore, there is a need to compare the effectiveness of both modes of training.

 This research presents a study that aims to investigate the effectiveness of digital and physical card games for refresher training of CHWs and contributes to the growing literature on the use of digital tools for healthcare training in resource-constrained settings, focusing on the usage of smartphone games for CHW refresher training in India.

The study involves four groups for a quasi-experiment. 
Firstly ones receiving the training through card game digitally  (n=92) (IG-1),
secondly ones receiving the training through card game physically  (n=92) (IG-2),
third ones receiving traditional classroom training  (n=92) (IG-3),
lastly ones with no training (n=92) (CG). Each group consists of 46 Accredited Social Health Activist (ASHAs) and 46 Anganwadi Workers (AWWs). Each intervention group is trained and evaluated for knowledge gain and retention, and the intervention groups are compared between them and the control group.

The research questions that the study is trying to answer are:
\begin{itemize}
    \item [\bfseries RQ1] Does knowledge gain and retention differ for participants taught using the traditional classroom method compared to playing digital and physical card games?
    \item [\bfseries RQ2] Is there a novelty effect of using games for training CHWs?
\end{itemize}

\section{Background}

\subsection{AAA workers}

In India, there are three cadres of Community Health Workers (CHWs):
\begin{itemize}
    \item Auxiliary Nurse-Midwife (ANM): ANMs provide healthcare services at sub-centers and visit villages. They have a broad set of responsibilities, including supporting AWWs and ASHA workers. ANMs work in a cluster of 4 to 5 villages and are responsible for community vaccination at a Sub-Centre level.
    \item Anganwadi Worker (AWW): AWWs are appointed for the Integrated Child Development Scheme (ICDS) project. They oversee mother and child nutrition, early childhood education (ECE) of children, and children's overall development, usually in their own community. They provide food supplements for children, adolescent girls, and lactating women.
    \item Accredited Social Health Activist (ASHA): ASHAs are community health mobilizers appointed by the MoHFW (Ministry of Health and Family Welfare) to facilitate the community in accessing health services. They usually work in their community promoting maternal and child health, immunizations, and institutional deliveries. They receive performance-based incentives to facilitate institutional delivery and immunization, provide basic medicines (like oral contraceptives), and refer critical patients to the community health subcenter.
\end{itemize} 
In India different cadres of CHWs have complementary and sometimes overlapping roles and responsibilities, collaboratively contributing towards improving maternal and child health outcomes.

\subsection{Related Works}

Previous research explored ways for effective communication about health training through radio for masses \cite{Kumar2015, Kumar2015a, Kumar2013, Yadav2017, Yadav2019, Yadav2019a, Ward2020}, timely notifications \cite{Perez2020}, and illustration of stepwise instructions for basic skills \cite{Tulaskar2020}. Researchers created informational video bytes for refresher training \cite{Ramachandran2010, Javaid2017}. Researchers conducted study on refresher training on the topic of anemia through quiz app on smartphone \cite{Majhi2021}. Further, researchers appended quiz with the video bytes, and incentivized listeners by providing talk time for cellular network if they provided correct answers \cite{Shah2017}. Researchers explored game based on Augmented Reality for collaborative play \cite{Majhi2022}, and support services \cite{Patel2019}. Researchers explored the potential of serious games for refresher training CHWs \cite{Majhi2024HCI_International} and collecting data on measuring efficiency of CHWs through geospatial games \cite{Majhi2024GoodIT}. None of the previous studies have not explored digital of physical card games to teach immunization for CHWs in a resource constraint settings.

In this research, the idea of cards as information blocks is to display vaccine information to players. Based on the information shown, the player makes decisions. Out of many physical tokens and objects of play, card gameplay was chosen as the game artifact. The possibilities of playing through card games were explored. Physical card games are ubiquitous. The primary reasons for this are its cost-effectiveness and handling ease. A deck of cards can be played in numerous different ways just by altering the rules. Different groups of cards can be added to the existing deck to enable new ways to play with it.  Through player engagement and collaboration, a significant learning advantage is observed \cite{Bochennek2007}. However, published research papers on the use of card games to train community healthcare workers were sparse.

The core learning objective of the game is for CHWs to memorize the sequence and dosage of immunizations and supplements required by the child and mother. Researchers \cite{Kappen2016} found that personalizing gaming experiences is important for older adults as they focus more on enjoyment than performance \cite{Birk2017}. Challenges specific to age often act as barriers that impact the relevance of gameful design elements for the elderly.

\subsection{Content for Refresher Training}
 Immunization cover was less than half in 163 Indian districts \cite{Panda2020}. Researchers \cite{Bag2017, Bashingwa2021} also reported that CHWs lack comprehensive knowledge of immunization schedules as part of the Mother and Child Protection Card (MCP card) given to every mother by the CHWs registering her pregnancy. Researchers \cite{Kizhatil2019} found that partial immunization in children strongly correlates with issuing MCP cards. The child immunization table from the MCP card was chosen as the primary content for the game. This schedule complies with the Indian Academics of Paediatrician (IAP) guidelines \cite{Kasi2021}. Also, Ante-Natal Care (ANC), and Post-Natal Care(PNC) services were included to complete the comprehensive 1000-days care information.

\section{Method}

\subsection{Aim}

The primary aim of this study is to investigate the use of digital tools, specifically training games, for enhancing the skills and knowledge of CHWs in resource-constrained settings. This study aims to provide information on designing, implementing, and evaluating a training game. In addition, we seek to explore the experiences of CHWs during gameplay and examine the game's impact on their knowledge.
CHWs are involved in the design and development process, ensuring that the game is tailored to their needs and preferences. Qualitative feedback from CHWs during the design process is essential to ensure the game is user-friendly, engaging, and relevant to their daily work. This research also contributes to the broader literature on digital tools for training healthcare workers and adds to the growing body of evidence supporting the use of participatory design approaches for developing training games.

\subsection{Participants}

The study involved an evaluation of the game by human participants, for which Institutional Review Board (IRB) approval was obtained to ensure ethical compliance with the Declaration of Helsinki \cite{DOH2013}. All participants were provided with both verbal and written information detailing the study's objectives, procedures, and expected outcomes. Written informed consent was obtained from all participants prior to their inclusion in the study.

The sample size was determined using G-power \cite{Erdfelder2007}. The calculations were based on an effect size (d) of 0.5, a power (1-\(\beta\)) of 0.95, and a significance level (\(\alpha\)) of 0.05. A medium effect size, suggests that the intervention is expected to have a moderate impact on the participants' performance, which is expected in behavioral research involving training and educational interventions. This effect size 0.5 indicates a meaningful difference between groups that is neither too small to be insignificant nor too large to be uncommon, aligning with the practical considerations of evaluating the game's effectiveness.

The calculation resulted in a required sample size of 88 participants per arm. Accounting for a potential 10\% attrition rate, it was estimated that 97 to 100 participants would be necessary for each of the four groups, with each group comprising 50 Accredited Social Health Activists (ASHAs) and 50 Anganwadi Workers (AWWs).

To recruit participants, CHW supervisors were contacted, and 200 ASHAs and 200 AWWs were selected through convenience sampling. By the conclusion of the study, 184 ASHAs and 184 AWWs remained, resulting in 46 ASHAs and 46 AWWs per group, totaling 92 participants per group. The attrition observed was primarily due to the unavailability of CHWs on the scheduled days for testing and evaluation, as they were required for emergency visits. Demographic details of the participants are provided in a Table in the supplementary materials.

\subsection{Participants}

The study involves the evaluation of the game by human participants, and Institutional Review Board (IRB) approval was obtained to ensure the ethical conduct of the study, in compliance with the Declaration of Helsinki \cite{DOH2013}. All participants received verbal and written information about the study's aim, procedure, and outcome and gave informed consent to participate in writing.

The participant selection process involved the following steps:

\begin{itemize}

    \item  Sample Size Calculation: G*Power \cite{Erdfelder2007} was used to calculate the sample size. An effect size (d) of 0.5 was chosen, which represents a medium effect size. In the context of this research, a medium effect size indicates that the intervention is expected to produce a noticeable and practically significant improvement in CHWs' knowledge and skills, but not an overwhelmingly large change. The power (1-\(\beta\)) was set at 0.95, ensuring a high probability of detecting a true effect if it exists, and the level of significance (\(\alpha\)) was set at 0.05, allowing for a 5\% chance of a Type I error. The calculated sample size was 88 participants for each arm. Considering a 10\% attrition rate, it was determined that 97-100 participants were required for each of the four groups.
    
    \item Participant Recruitment: CHW supervisors were contacted to facilitate the recruitment process. A total of 200 Accredited Social Health Activists (ASHAs) and 200 Anganwadi Workers (AWWs) were selected through convenience sampling, amounting to 100 participants (50 ASHAs and 50 AWWs) for each of the four groups.
    
    \item Informed Consent: All participants received detailed information about the study and provided written informed consent before participating.
    
    \item Final Sample Size: By the end of the study, 184 ASHAs and 184 AWWs were retained, resulting in 92 participants per group (46 ASHAs and 46 AWWs for each group). The attrition was primarily due to the unavailability of CHWs on the days of testing and evaluation, often due to emergency visits. 
    
    The demographic details of the participants are provided in the Table appended in the Supplementary Materials.
    
\end{itemize}

\subsection{Apparatus}

\subsubsection{Physical Card Play as a Learning Tool}

\subsubsection{Designing the cards}

Content was categorized into four silos depending on the time it takes to provide care to stakeholders. Four silos of cards were made, each symbolizing an immunization, medicine supplement, or care activity or an event. They are classified as follows: children under one year (n = 22), children above one year (n = 15), Ante-Natal Care (ANC) (n = 14), and Post-Natal Care(PNC) (n = 9) combined to form a deck of 60 cards. These playing cards (5.5 X 8.5 cm) are similar in size to credit cards but made of card paper (200 GSM). 

A collection of 60 cards is appended in the Supplementary Materials.

\subsubsection{Framing Challenges and Rules}

Before commencing gameplay, players should shuffle a deck of 60 cards and evenly distribute them among four players, with each receiving 15 cards. During their turn, players must place a card from one silo onto the corresponding placeholder on the board. Cards are numbered sequentially by month and year, with additional fractional numbering indicating multiples within an age group. This numbering system aids players in tracking card placement progress within a sequence. Successfully placing a card grants the player two additional turns as a bonus. If no valid card placements remain or if two bonus rounds have been exhausted, play passes to the next player. Subsequent players may then place cards in age groups either higher or lower than the current set on the board. Gameplay continues until all cards have been played. The player to exhaust their hand first wins the round, earning the top rank, followed by remaining players until all cards are played out.

\subsection{Experimental Design}

The study follows a quasi-experimental design (QED) because a true experimental design (TED) would not be possible to conduct in this context. The same location and relatively small sample make it difficult to randomize  \cite{Harris2006}. Thus, QEDs have lower internal validity than TEDs. However, QEDs have higher external validity because the intervention is followed by measurement of the outcome \cite{Harris2006}. In this QED, the game's design is used to evaluate the benefits of the intervention.

\subsection{Evaluation}
CHWs were evaluated thrice in different phases using pen-and-paper questionnaire surveys. The questionnaire starts with a consent form and space to provide demographic details, followed by 40 single-answer, multiple-choice questions of almost equal weightage. The CHWs are supposed to tick a checkbox against the right answer, indicating their choice of option. The 3 Page Questionnaire is appended in the Supplementary Materials.

A baseline survey was conducted to check the existing knowledge of the CHWs. Then, the intervention was conducted for the intervention groups. All four groups formed a pair of AWW as a team and ASHA as another team (2X2 team formation - Total 4 players) for every round. IG1 was conducted through card games on their smartphones. IG2 was conducted through a physical or paper card deck. IG3 has imparted the same knowledge through in-person classroom training. Similar surveys were conducted for CG, but no training was provided to check for placebo effects.

After the intervention or three rounds of gameplay, a post-test was conducted to check the knowledge gains of the CHWs. We should consider that there might be a potential novelty effect or the chances of being effective when initially adopted but fade with time. Therefore, it becomes crucial to understand the long-term effects of the intervention on learning through longitudinal research. After three weeks, a delayed retention test (DRT), similar to the post-test, was performed for all players. According to previous literature \cite{Haynie1994,Nungester1982} on DRT, a 3-week delay length (more than or equal to 2) was chosen to represent a typical span for DRT. Players were not briefed or previously informed about conducting DRT. CG was not trained throughout the study. Later, after the study, CG was also trained by master trainer as well as smartphone game based training to match the benefits of IGs and bridge the knowledge gap.

\section{Findings}

The scores for 3 phases of the tests for the four groups are tabulated in a Table, appended in the Supplementary Materials. After testing assumptions for t-tests like normality and sphericity for all four group scores, Pairwise t-tests were performed to check within-group and between-group differences at every phase of the evaluation. Repeated Measure ANOVA (RM-ANOVA) was also performed to confirm whether an overall difference in scores was found in each phase of the evaluation.

IG-1 and IG-2, or both Card Game groups, performed significantly better than IG3 (Classroom group) and CG(Control Group) (p<0.05), showing the effectiveness of teaching through card games. Though the IG1 or digital games group scored more than the IG2 or physical game group, significant differences were not observed between IG1 and IG2 (p>0.05), signifying a marginal difference in effectiveness between the two gameplay modes. The CG or control group performed significantly worse than all other groups, nullifying the placebo effects (p<0.05). This confirms the RQ2.
The long post-test scores for IG1 and IG2 were almost similar, showing a decline in post-test scores. The Control group also showed similar trends. This signifies that the knowledge refresher efforts fade without reinforcement learning.

\section{Discussion}

In summary, the playful activity with immunization cards was significantly effective for refresher training of CHWs with significant knowledge retention. The results were compared by repeating the experiment with both groups. The digital card mode was significantly better than the physical card mode, traditional in-person classroom mode, and control group, except in a few cases.

\subsubsection{Trends:}
A subtle trend shows that the delta change in knowledge gains between pre- and post-tests decreased with age. However, the trend was insignificant in this investigation, with a confidence interval 95\%. Research on learning in other contexts also finds similar results. Years of formal education also have an insignificant effect on the baseline score, as the knowledge or experience gained depends mainly on the years of experience, working in the field as an ASHA worker, and having hands-on experience dealing with common issues. As many ASHAs had 9-10 years or more of experience, a group comparison between them and the other minority would not be statistically significant.

\subsubsection{Inconsistency in knowledge:}
While checking the questionnaire survey, many mismatches were found regarding the standardization of information. Many CHWs ignored two questions on Rotavirus. Further investigation found that the Rotavirus vaccines were not operational in Madhya Pradesh and, thus, were not included in the printed brochures. Some of the ASHAs made mistakes in the PCV booster. Further investigation found that the term "booster" was not mentioned for any of the vaccines in the schedule they were trained with, thus creating confusion. But most of them understood and answered it correctly. 

\subsubsection{Malpractices:}
During the baseline survey, three CHWs were caught in the malpractice of cheating from their job aid or immunization schedule booklet. They were excluded from the statistical evaluation calculations. The study design used the same questions for all three survey rounds. In the next phase, the order of questions was randomized to minimize the question's positional effect and order remembrance. Three sets of questionnaires were made for each language with a randomized order of questions.

\subsubsection{Language as a barrier:}
Some ASHAs, especially those with lower formal education, struggled with reading Hindi numbers printed on the cards while completing the survey questionnaire. During design iterations, the numbers were changed to English, and the words in Hindi remained unchanged. Some ASHAs studied in a Madrasa and could only recognize English and Urdu numbers. They required translation of the numbers in English to fill out the survey form but still managed to play cards as the rest of the information on the card was enough to play. We changed all numbers in the card and the app to English numbers. This didn't affect the results as we discarded the previous ones from the calculations.

\subsubsection{Focused Group Discussions:}
FGDs with CHWs reveal much information about their lives and daily activities, which usually gets masked behind their work's top-down bureaucratic framework. Some ASHAs reported that the ANM is responsible for immunization, and the ASHA is mostly responsible for community mobilization, coordination, and assistance with the ANM. Despite that, any mistake the ANM commits, like wrong or missed immunization or not practicing hygiene standards, is often blamed on the vulnerable position of the lower level, which in this case happens to be an ASHA. Still, an ASHA does her best to provide the assistance the ANM and the community require. If CHWs could be trained with frequent refresher training, they could help minimize these errors in their daily job to a large extent, thus strengthening community healthcare services. Most ASHAs are intrinsically motivated to learn about healthcare and provide better care for the community. Despite the selfless efforts the CHWs put in, they were rarely appreciated by the state and sometimes by the community. ASHAs of India are recognized as one of the six awardees of the Global Health Leaders Awards at the 75th World Health Assembly \cite{Asthana2022} for their important role in bridging the community and those living in rural poverty to access primary healthcare services during the crucial times of the COVID-19 pandemic.

\subsubsection{Play-testing experiences}

Familiarity with card games in general, facilitates the understanding of the rules of play. After playing 2-3 rounds, the CHWs tended to memorize the sequence and didn't need to go back to the immunization tables for reference. After shadowing CHWs for 2-3 weeks, we found they eagerly formed teams to play and have fun by challenging their opponents. Sometimes, during the gameplay, a team starts criticizing the knowledge of the opponents, which might demotivate the opponents and hinder learning outcomes.

\subsubsection{Limitations of the study}

The experiment was carried out in the formal institutional settings. CHW cadre serves the state and is obliged to function within the framework of the top-down bureaucracy. This might have limited how they interact in a team's collaborative game setup or freely given feedback to the researchers.

\section{Conclusion}

This research investigates the effects of the collaboration of two cadres of CHWs (ASHAs and AWWs) in team-up gameplay. The study compares whether knowledge gain and retention differ for participants taught using the traditional classroom method compared to playing digital and physical card games. We found that there were significant differences in knowledge gain between them (RQ1). Digital and physical game groups outperformed the classroom training group and control group. However, knowledge retention is significantly better in digital or smartphone gameplay. We found that pairing 2 ASHA and 2 AWW as a group in gameplay produced statistically significant results compared to the previous studies. We didn't find placebo effect of not using games (RQ2).

Our contribution lies in (1) [Literature] The research broadens the literature on effectiveness and impacts of physical and digital games as tools for refresher training healthcare workers from two cadres playing collaboratively. (2) [Tool/Artifact] Designing a deck of cards on the topic of immunization, which can be used as an artifact for building new games out of new playing rules, (3) [Survey Form] Creating a 3-page Questionnaire to evaluate knowledge gained through the gameplay, (4) [Software/Tool]  Creating an Android game designed in Unity, hosted as free open access for downloading from Google and Apple PlayStores (5) [Repository] Open Access Unity Project, code and materials for forking/modifying on GitHub.

\section{Future Work}

This study compares the effectiveness of immediate learning and long-term retention but does not examine the learning process. Upcoming studies might focus on how participants learn in both conditions to understand the causes of differences in learning and motivation.

\begin{acks}

We wish to express our deep appreciation to all the Community Health Workers (CHWs) and their supervisors for their active participation, support, and encouragement of their colleagues in this study. We are also grateful to the NGOs, their teams, and the associated organizations for their essential contributions to this research. We acknowledge the reviewers for their valuable and constructive feedback. Finally, we extend our sincere thanks to the Science \& Engineering Research Board (SERB), the Federation of Indian Chambers of Commerce \& Industry (FICCI), and UNICEF, New Delhi, for their generous support of our research.

\end{acks}

\bibliographystyle{ACM-Reference-Format}
\bibliography{references}

\end{document}